\documentclass[a4paper]{jpconf}

\usepackage{amsmath}
\usepackage{graphicx}
\usepackage{subfigure}

\begin{document}
\title{Search for New Physics in Rare Top Decays}

\author{Pratishruti Saha\footnote{Based on 
K.~Kiers \textit{et al.} Phys.\ Rev.\ D {\bf 90}, 094015 (2014) and 
P.~Saha \textit{et al.} Phys.\ Rev.\ D {\bf 90}, 094016 (2014).}}

\address{Physique des Particules, Universit\'{e} de Montr\'{e}al,
         C.P. 6128, succ centre-ville, Montr\'{e}al, QC, Canada H3C 3J7}

\ead{pratishruti.saha@umontreal.ca}

\begin{abstract}
Top physics provides a fertile ground for new-physics searches. At present, 
most top observables appear to be in good agreement with the respective Standard Model 
predictions. However, in the case of decay modes that are suppressed in the Standard Model, 
new-physics contributions of comparable magnitude may exist and yet go unnoticed 
because their impact on the total decay width is small. Hence it is interesting to probe rare top decays. 
This analysis focuses on the decay $t \to b \bar b c$. Useful observables are identified and prospects for 
measuring new-physics parameters are examined.
\end{abstract}

\section{Introduction}
The top quark was discovered at the Tevatron in 1995~\cite{top_discovery_CDF,top_discovery_D0}. 
Since then, many of its properties have been measured at the Tevatron and the Large Hadron Collider (LHC). 
Almost all of these seem to agree rather well with the Standard Model~\cite{top_CDF,top_D0,top_ATLAS,top_CMS}. 
For a while, the forward-backward asymmetry in $t \bar t$ production at the Tevatron seemed to be much larger 
than predicted in the Standard Model~\cite{afb_older}. However, in recent analyses, the discrepancy seems 
to have decreased considerably~\cite{afb_D0}. Moreover, at the LHC, no deviation was observed at any stage 
in the analogous charge asymmetry observables~\cite{charge_asymm_LHC_ATLAS,charge_asymm_LHC_CMS}. 
At present, the top quark does seem to behave exactly as described in the Standard Model. 
Nonetheless, the LHC, being a top-factory, affords us the opportunity to probe top properties further 
and also study rare processes involving the top. One such process is the decay $t \to b \bar b c$.

The dominant decay modes of the top involve $t \to W^+ b$, with the $W^+$ then decaying to 
$l^+ \nu$, $u \bar d$ or $c \bar s$. The decay $t \to b \bar b c$ would arise if $W^+ \to c \bar b$.
However, this decay is extremely suppressed in the Standard Model as it is proportional 
to $V_{cb}^2$ ($V_{cb}$ = 0.04 being one of the elements of the Cabibbo-Kobayashi-Maskawa matrix).
New physics contributions to this decay, if present, can be parametrized in terms of the following
effective Lagrangian :
\begin{equation}
{\cal L}_{eff} \quad = \quad {\cal L}_{eff}^V \;+\; {\cal L}_{eff}^S \;+\; {\cal L}_{eff}^T
\end{equation}
where
\begin{align}
{\cal L}_{eff}^V 
\; &= \; 
4\sqrt{2}G_F V_{cb}V_{tb} \,
\big\{ \quad\;
X_{LL}^V \; (\bar b \gamma_\mu P_L t) \; (\bar c \gamma^\mu P_L b) 
\; + \;
X_{LR}^V \; (\bar b \gamma_\mu P_L t) \; (\bar c \gamma^\mu P_R b) \nonumber \\ 
&\qquad\qquad\qquad\;\;\; 
\; + \; 
X_{RL}^V \; (\bar b \gamma_\mu P_R t) \; (\bar c \gamma^\mu P_L b)
\; + \;
X_{RR}^V \; (\bar b \gamma_\mu P_R t) \; (\bar c \gamma^\mu P_R b) \;
\big \}
\; + \; h.c. 
\displaybreak \\
{\cal L}_{eff}^S 
\; &= \; 
4\sqrt{2}G_F V_{cb}V_{tb} \,
\big\{ \quad\;
 X_{LL}^S \; (\bar b P_L t) \; (\bar c P_L b)
\; + \;
X_{LR}^S \; (\bar b P_L t) \; (\bar c P_R b) \nonumber \\
&\qquad\qquad\qquad\;\;\; 
\; + \;
X_{RL}^S \; (\bar b P_R t) \; (\bar c P_L b)
\; + \;
X_{RR}^S \; (\bar b P_R t) \; (\bar c P_R b) \, 
\big\}
\; + \; h.c. \\[2ex]
{\cal L}_{eff}^T 
\; &= \; 
4\sqrt{2}G_F V_{cb}V_{tb}
\big\{
X^T_{LL} \; (\bar b \sigma^{\mu\nu} P_L t) \; (\bar c \sigma_{\mu\nu} P_L b) 
\; + \;
X^T_{RR} \; (\bar b \sigma^{\mu\nu} P_R t) \; (\bar c \sigma_{\mu\nu} P_R b) \, 
\big \}
\; + \; h.c.
\end{align}
In the above, the colour indices have been suppressed but they are assumed to contract in the same manner as in the 
Standard Model.

At the LHC, the largest number of top quarks are produced through the process $pp \to t \bar t$.
The decays of these tops and antitops can then be investigated for signs of new physics.
The large production cross-section compensates for the small branching fraction such that  
a statistical analysis remains feasible. Moreover, as shown in Refs.~\cite{paper_1} and \cite{paper_2},
angular correlations between $t$ and $\bar t$ decay products also carry the imprint of new physics.

\section{Useful Observables}

For the purposes of this analysis, it is assumed that the $t$ decays to $b$, $\bar b$ and $c$, while
the $\bar t$ decays to a $\bar b$, a charged lepton and an antineutrino.
The square of the matrix element for the process $g g \to t \, \bar t \, \to b \, \bar b \, c \, \bar b \, \ell^- \, \bar \nu_{\ell}$
is calculated in Ref.~\cite{paper_1}. It is seen that all new-physics dependence in $|{\cal M}|^2$ can be 
parametrized in terms of certain combinations $\hat A^{\sigma}_k$ of the couplings $X^I_{AB}$ given by 
\begin{align*}
\hat A_{\bar b}^+ 
\quad &= \quad 
4 \left|X^{V}_{LL}\right|^2 - 8 \,\mbox{Re}\left(X^T_{LL}X^{S*}_{LL}\right)+32 \left|X^T_{LL}\right|^2 \, ,\\[0.5ex]
\hat{A}_{\bar b}^- 
\quad &= \quad
4\left|X^{V}_{RR}\right|^2 - 8 \,\mbox{Re}\left(X^T_{RR}X^{S*}_{RR}\right)+32 \left|X^T_{RR}\right|^2 \, ,\\[0.5ex]
\hat{A}_{b}^+ 
\quad &= \quad
\left|X^{S}_{LL}\right|^2+\left|X^{S}_{LR}\right|^2 -16\left|X^{T}_{LL}\right|^2 \, ,\\[0.5ex]
\hat{A}_{b}^- 
\quad &= \quad 
\left|X^{S}_{RR}\right|^2+\left|X^{S}_{RL}\right|^2 -16\left|X^{T}_{RR}\right|^2 \, ,\\[0.5ex]
\hat{A}_{c}^+ 
\quad &= \quad 
4\left|X^{V}_{LR}\right|^2 + 8 \,\mbox{Re}\left(X^T_{LL}X^{S*}_{LL}\right)+32\left|X^T_{LL}\right|^2 \, ,\\[0.5ex]
\hat{A}_{c}^- 
\quad &= \quad
4\left|X^{V}_{RL}\right|^2 + 8 \,\mbox{Re}\left(X^T_{RR}X^{S*}_{RR}\right)+32\left|X^T_{RR}\right|^2 \, .
\end{align*}
In fact, as discussed in Ref.~\cite{paper_2} (Sec. III B 3), this is true even for 
$q \bar q \to t \, \bar t \, \to b \, \bar b \, c \, \bar b \, \ell^- \, \bar \nu_{\ell}$.

If the $t$ decay contains contributions from new physics, then, several observables related to the 
decay products get affected. Of these, two turn out to be specially useful. These are
\begin{enumerate}
 \item $\dfrac{d\sigma}{d\zeta^2_{ij}}$ where $\zeta^2_{ij} = \dfrac{(p_i + p_j)^2}{m_t^2}$ and  
       $p_i$ and $p_j$ are the momenta of two out of the three particles coming from the top decay. 
 
 \item $\dfrac{d^2\sigma}{d\cos\theta^*_i d\cos\theta^*_{\ell}}$ where $\theta^*_i$ is the angle between 
       the direction of the $i^{th}$ decay product of the $t$ and the spin quantization axis of the $t$ 
       in the $t$ rest frame, and $\theta^*_{\ell}$ is the angle between the direction of the $\ell^-$ and 
       the spin quantization axis of the $\bar t$ in the $\bar t$ rest frame.
\end{enumerate}

\begin{figure}[!htbp]
\centering
\subfigure[]
{
\includegraphics[scale=0.8]{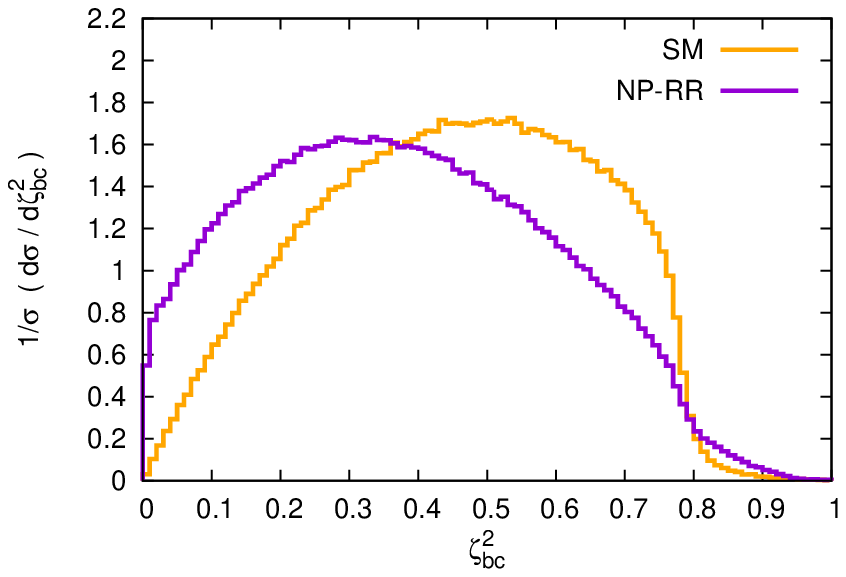}
\label{fig:zetasq_bc}
}
\subfigure[]
{
\includegraphics[scale=0.8]{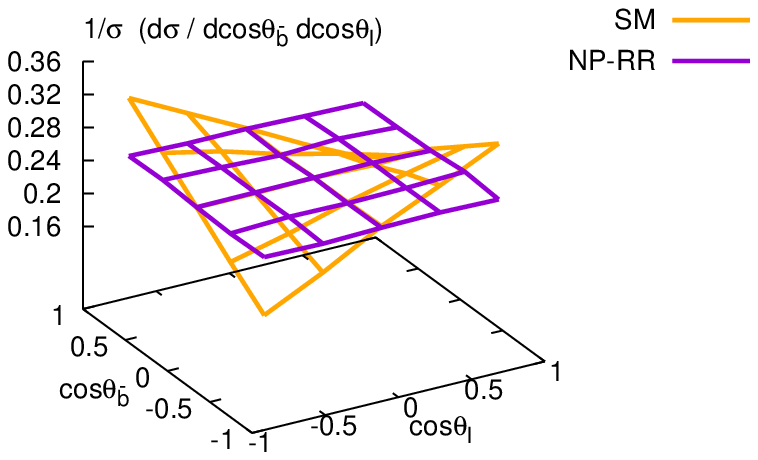}
\label{fig:ctheta_bbarl}
}
\caption{Normalised distributions for (a) ${d\sigma}/{d\zeta^2_{bc}}$ and (b) 
${d^2\sigma}/{d\cos\theta^*_{\bar b} d\cos\theta^*_{\ell}}$.
NP-RR represents the new physics scenario where all $X^I_{RR} = 1$ and all 
$X^I_{LL} = X^I_{LR} = X^I_{RL} = 0$.}
\label{fig:observables}
\end{figure}

The detailed expressions for these for different choices of $i$ and $j$ can be found in Ref.~\cite{paper_1}.
Fig.~\ref{fig:observables} shows the normalised distributions for ${d\sigma}/{d\zeta^2_{bc}}$
and ${d^2\sigma}/{d\cos\theta^*_{\bar b} d\cos\theta^*_{\ell}}$ for a particular choice of new-physics couplings.
It clearly demonstrates the discriminating power of these observables even for cases where the new-physics couplings 
are not large.

\pagebreak

That is not all. ${d\sigma}/{d\zeta^2_{ij}}$ and ${d^2\sigma}/{d\cos\theta^*_i d\cos\theta^*_{\ell}}$
form a set of six differential distributions, each dependent on a different combination\footnote{In the case
of ${d\sigma}/{d\zeta^2_{\bar b c}}$, there is also some additional dependence on Re($X^V_{LL}$).} 
of the six $\hat A^{\sigma}_k$'s.
By performing a simultaneous fit to all six distributions, it is possible to extract the values of these 
$\hat A^{\sigma}_k$'s. Ref.~\cite{paper_2} discusses the detailed procedure for carrying out this 
exercise. In the absence of actual data in this channel, Monte Carlo data generated using 
MadGraph 5~\cite{Madgraph} is used to test and establish the algorithm. The model files 
for MadGraph have, in turn, been generated using FeynRules~\cite{FeynRules}. The crux of the idea 
is as follows. Each of ${d\sigma}/{d\zeta^2_{ij}}$ and 
${d^2\sigma}/{d\cos\theta^*_i d\cos\theta^*_{\ell}}$ gives a histogram.
In general, each histogram $\mathtt{H}$ can be written as 
\begin{equation}
\mathtt{H = a_0 T_0 + a_1 T_1 + a_2 T_2 + a_3 T_3 + a_4 T_4 + a_5 T_5 + a_6 T_6 + a_7 T_7}
\end{equation}
where $\mathtt{T_0}$ is the histogram $\mathtt{H}$ for the Standard Model and other $\mathtt{T_n}$'s represent 
same histogram $\mathtt{H}$ with $\hat A^{\sigma}_k$ and Re($X^V_{LL}$) = $\mathtt{nonzero}$, one at a time.
In essence, one only needs to determine the weights $\mathtt{a_n}$ that best reproduce the data.

\section{Role of $t \bar t $ spin correlations}

The distributions ${d^2\sigma}/{d\cos\theta^*_i d\cos\theta^*_{\ell}}$ are, in fact, closely related to the 
$t \bar t $ spin correlations. In the Standard Model,
\begin{equation}
\dfrac{1}{\sigma} \; \dfrac{d^2\sigma}{d\cos\theta^*_i \; d\cos\theta^*_{\ell}} \;\;=\;\;
\dfrac{1}{4} \left( 1 + \kappa_{t \bar t} \, \alpha_i \, \alpha_{\ell} \, \cos\theta^*_i \, \cos\theta^*_{\ell} \right) \, ,
\end{equation}
where $\kappa_{t \bar t}$ is the spin correlation coefficient and the factors $\alpha_i$ and $\alpha_{\ell}$
depend on the identity of the particles $i$ and $\ell$. Since the $t \bar t$ pair decays instantly, the spin correlation
coefficient is inferred by fitting this distribution to the functional form $c_1 + c_2 \, \cos\theta^*_i \, \cos\theta^*_{\ell}$.
The coefficient $c_2$ contains $\kappa_{t \bar t}$.
In the presence of new physics, this distribution gets modified to the form
\begin{equation}
\dfrac{1}{\sigma} \; \dfrac{d^2\sigma}{d\cos\theta^*_i \; d\cos\theta^*_{\ell}} \;\;=\;\;
\dfrac{1}{4} \left( 1 + \kappa_{t \bar t} \, \alpha_i \, \alpha_{\ell} \, (1 + \beta_i) \,\cos\theta^*_i \, \cos\theta^*_{\ell} \right) \, ,
\end{equation}
where $\beta_i$ is a different function of the $\hat A^{\sigma}_k$'s for $i = b, \bar b \text{ and } c$.
\pagebreak
If the value of the spin correlation coefficient were to be inferred from this now, it would yield
$\kappa_{t \bar t}^{\prime} = \kappa_{t \bar t} \, (1 + \beta_i)$.
There is, however, a caveat in all of this. Note that the final state contains two $\bar b$'s.
In order to construct the aforementioned observables, the $\bar b$ coming from the $t$-decay has to be `correctly'
identified. Here and in Ref.~\cite{paper_2}, the $\bar b$ that yields the smaller value of 
$\Delta m = |m_t - \sqrt{(p_b + p_{\bar b} + p_c)^2}|$ is taken to be the one arising from the $t$-decay.
If both $\bar b$'s yield nearly identical values of $\Delta m$, the event is discarded. This `cut' modifies 
the shape of the angular correlations such that it is no longer possible to fit them to the functional form 
$c_1 + c_2 \, \cos\theta^*_i \, \cos\theta^*_{\ell}$ or any simple modification of it.
Nonetheless, the histogram fitting described in the previous section works just as well since the same `cut' 
is applied to the data ($\mathtt{H}$) as well as the templates ($\mathtt{T_n}$).

\section{Results and Conclusions}

A full fit to all six observables is performed using Monte Carlo data with statistics corresponding to 
an integrated luminosity $\sim$ 300 fb$^{-1}$ for 14 TeV collisions at the LHC.
Different new-physics scenarios corresponding to different choices of the $X^I_{AB}$'s are examined. 
The values of $\hat A^{\sigma}_k$ are recovered to within $\pm$ 1.7$\sigma$ of their input values.
If only the $\zeta^2_{ij}$ distributions are measured, the values of ($\hat A_{b}^+ + \hat A_{b}^-$),
($\hat A_{\bar b}^+ + \hat A_{\bar b}^-$) and ($\hat A_{c}^+ + \hat A_{c}^-$) 
can still be determined.

\end{document}